# Experimental Evidence for Zero DC Resistance of Superconductors

S. Sarangi, S.P. Chockalingam, Raghav G. Mavinkurve and S.V.Bhat*

Department of Physics, Indian Institute of Science, Bangalore 560012, India

**Abstract**

Even after nearly a century of discovery of superconductivity, there has been no direct experimental proof of the expected zero resistance of superconductors. Indeed, it has been believed that it is impossible to experimentally show that the resistance has fallen exactly to zero. In this work we demonstrate that the dc resistivity of a superconducting material below the transition temperature has to be exactly zero.

Perfect conductivity or zero resistance is one of the two defining properties of superconductivity, the other being perfect diamagnetism [1]. However, the question of whether the resistance of a superconducting material to direct current flow below the superconducting transition temperature $T_c$ is identically zero or only unmeasurably small is not yet settled experimentally [2]. Using the method of decay of persistent currents in superconducting rings the upper limit of resistivity in superconductors has been variously estimated as 2 x 10 $^{-18}$ Ω-cm [3] or 7 x 10 $^{-23}$ Ω-cm [4], which is undoubtedly a very small value. Yet it is important to have an unambiguous determination of the resistance to decide if it is indeed zero. In this letter using a simple experimental arrangement we demonstrate that the resistivity of a superconducting material below $T_c$ has to be absolutely zero.

In the classic experiment [5] of Kammerlingh Onnes the resistance of the mercury sample precipitously dropped from 0.08 Ω above 4 K to below the sensitivity limit of his experiment ( ~ 3 x 10 $^{-6}$ Ω ) when cooled to about 3 K. For this experiment Onnes used the conventional method of determining the resistance, namely, the measurement of voltage drop across the sample when a known current is passed through it. The sensitivity of the apparatus and therefore the smallest resistance that can be measured using this technique has improved significantly over the decades. Yet, it is still not enough to determine the very low resistance, if any, of the superconductors. Moreover, the technique also suffers from problems such as thermal fluctuations, thermoelectric power in inhomogeneous materials, and the transition resistance of the potential leads [4]. Along with the advent of the high-$T_c$ superconductivity [6-11], the four-probe method of measuring the resistance has almost universally been adopted. This method eliminates the problem of the contact resistance of the leads, which is particularly severe during the measurement of low resistances. However the ultimate sensitivity of the measuring voltmeter and the ammeter limits the lowest value of the resistance that can be determined even using this technique and sets an upper limit to it. For example, in their Nobel-prize winning work on the Ba-La-Cu-O system, Bednorz and Muller observed [6] a sharp (>90%) drop of resistance and inferred superconductivity. Wu et al. [7], concluded that the Y-Ba-Cu-O compound attained “zero-resistance” when the sensitivity of their apparatus could only set an upper limit: $\rho < 3 \times 10^{-8}\ \Omega$-cm. Sheng and Hermann[8] observed a decrease to $< 10^{-8}\ \Omega$-cm of the resistance of their $Tl_2Ba_2Cu_3O_{8+x}$ sample and concluded that they found strong evidence for the existence of high Tc superconductivity in the sample. Even with the most sensitive method of measuring the resistivity so far, namely, SQUID picovoltometry, an upper limit of $\sim 3 \times 10^{-11}\ \Omega$-cm could be set [12].

An extremely sensitive method for the purpose, pioneered by Onnes himself, is the technique of estimating the upper limit of the resistivity by studying the decay rate of the persistent current in a superconducting ring. Once established, the time dependence of the current I(t) through the ring is given by $I(t) = I_0\, e^{-(R/L)\, t}$ where $I_0$ is the current at t = 0, R is the resistance and L is the inductance of the ring. If the superconductor had zero resistance, the current would not decay even for infinitely long times. However, an experiment can be performed only over a limited amount of time. In a number of such experiments no detectable decay of the current was found for periods of time extending to several years. These experiments were used to establish lower limits on the lifetime of the current and from that the upper limits on the resistivity of the superconductor were calculated. Using this technique and taking in to account the possibility of the decay in type II superconductors due to effects like flux creep [13], for conventional superconductors it was estimated that $\rho < 10^{-23}\ \Omega$ cm [14-16]and for copper oxide superconductors $\rho < 10^{-18}\ \Omega$ cm [3] and $\rho < 10^{-22}\ \Omega$ cm [4]. While these upper limits are quite low, it is all the same believed [2] that, "… it is fundamentally impossible to demonstrate in an experiment the assertion that the resistance has fallen to exactly zero. An experiment can only ever deliver an upper limit for the resistance of a superconductor".

The quantitative determination of the resistance will invariably depend on the sensitivity of the measuring instruments and therefore it is conceivable that one can never 'measure' 'zero resistance'. In the following, we describe the results of our experiments, whose conclusions are based on qualitative comparison and therefore are free from the constraints of the two approaches adopted so far, and conclusively establish that the dc resistivity of a superconductor is indeed zero. Of course, it is well established and understood that the resistance below Tc need not be zero for ac transport and/or in the presence of an applied magnetic field. The resistive response of the normal electrons/quasiparticles, which is shunted out by the super-electron pairs as far as dc transport is concerned, is non-zero for finite frequencies. Similarly, the motion of quantized flux lines in the mixed state of a type II superconductor could also lead to dissipation. Thus 'zero resistance' of a superconductor is discussed in the context of zero magnetic field and zero frequency of the applied electric fileld.

The experimental arrangement used by us is sketched in fig.1. A and B form the two parallel arms of a superconducting loop of NbTi superconducting wire. A is a straight wire of length $L_A$ (about 8 cm in one of our experiments) and B, of length $L_{B,}$ (~300 cm in our experiments) is in the form of a circular coil in the proximity of which a Hall probe sensor S is placed. The principle of the experiment is as follows: to start with, path A is disconnected and a current **I** (= 6 Amperes in a typical experiment) from a constant current source is passed through path B. This generates a magnetic field $H_{n1}$ whose value ( = 174 Gauss) is measured by a Hall probe Gauss meter. Then the path A is connected in parallel with B, (While doing this respective ends of the two paths are twisted together for about 2.5 cm each so that a single NbTi path connects the NbTi loop to the copper wire on either side. This is to eliminate the effects of any possible asymmetric bifurcation of the current and the consequent differential terminal resistance when the loop goes superconducting.) and the field $H_{n2}$ generated by B is again measured. It is found that $H_{n2}$ = 4.5 Gauss which is nothing but $H_{n1}/(k+1)$, where $k = L_B / L_A$ is the ratio of the two lengths. This is a simple consequence of the fact that the resistance in the normal state of segment A, $R_{nA}$ , is k times less than $R_{nB}$ leading to $\mathbf{I}_{nB} = \mathbf{I}_{nA}/k$ and the total current $\mathbf{I} = \mathbf{I}_{nB}(k+1)$.

Now the assembly is cooled to liquid helium temperature (4.2 K), i.e., below the transition temperature of the NbTi wire ($T_c$= 9.3 K). In principle two scenarios are possible depending upon whether the resistance of the wire in the superconducting state is unmeasurably small but finite or indeed zero. First let us examine the result expected if the resistivity is extremely small, say on the order of $10^{-20}$ Ω-cm. (As we shall see, the actual magnitude is immaterial to our conclusion.) Then for an infinitesimally small value of resistivity ε, the resistance of segment A, $r_{sA} = \varepsilon L_A$ times the area of cross section of the wire and that of B, $r_{sB} = \varepsilon L_B$ times the area of cross section, such that $r_{sB} = k\, r_{sA}$. Consequently, $I_{sB} = I_{sA}/k$ and the magnetic field produced by coil B as measured by the Gauss meter, $H_{s2,}$ would be equal to $H_{n2}$.

If, instead, the resistance in the superconducting state is zero, both the paths are equal as far as their resistances are concerned and therefore carry equal currents (= **I**/2 ). As a result the field generated by coil B will be $H_{n1}/2$. In fig.2 (a) we present the magnetic field generated by and the current passing through B as a function of temperature for a typical value of the total current of 6 Amperes. It is clearly seen that in the superconducting state the magnetic field produced $H_{s2}$ is equal to 87 Gauss, i.e. equal to $H_{n1}/2$ and the current passing through the coil B is 3 Amperes i.e. exactly half of the total current. This result is possible only if the resistance of both the paths are equal and therefore identically zero. If on the other hand there is a finite resistance, irrespective of how small its magnitude is, even in the superconducting state the two paths will have to have the same ratio of their resistances as in the normal state. Then the current in the path B will be k times smaller than the current in path A and the field produced would be much less (in fact, the same as that in the normal state) than what we have observed. This expectation was verified by repeating the experiment with nearly identical configuration of the two paths but with copper instead of NbTi wire and the result is summarized in fig. 2(b). In this case, it is observed that only a small fraction of the current continues to flow even down to the lowest temperature in path B.

In a minor variation of the experiment, after the loop became superconducting, the source current was switched off, the superconducting loop being driven into the persistent current mode. It was observed that even now the field generated by coil B remained much larger than the value in the normal state, indicating that the resistances in the two paths are exactly zero. This provides additional evidence that no extraneous effects such as differential terminal resistances have any role to play.

In summary, we have demonstrated that the dc resistance of a superconducting wire is indeed zero and not just unmeasurably small, thus resolving the uncertainty that had lingered on for nearly a century after the discovery of the phenomenon of superconductivity.

The funding provided by the University Grants Commission, India for this work is gratefully acknowledged.

-------------------------------

*Electronic address for correspondence: svbhat@physics.iisc.ernet.in.

**FIGURE LEGENDS:**

**Figure 1:** The sketch of the experimental set up used. The part of the assembly enclosed in the dashed box could be inserted in an Oxford Instruments continuous flow cryostat for temperature variation. The coil B has 50 turns of NbTi wire of diameter 18 mm. Path A is made up of about 8 cm of straight length of the same NbTi wire. Respective ends of the two paths are twisted together for about 2.5 cm each so that a single NbTi path connects the NbTi loop to the copper wire on either side. This is to eliminate any possible effects of asymmetric bifurcation of the current and differential terminal resistance between the superconducting loop and the normal copper wire.

**Figure 2:** **(a)** Temperature dependence of the magnetic field H, sensed at the bottom of coil B. The total current used is 6 Amperes. When path A is not connected in parallel with B, all the 6 Amperes flow through B and produce a magnetic field $H_{n1}$ = 174 Gauss. When A and B are connected in parallel, a current of only 0.156 Ampere flows through path B and produces a magnetic field of 4.5 Gausss in the normal state, (T > 9. 3 K). However, in the superconducting state, the field produced by the coil B is measured to be 87 Gauss, corresponding to a current flow of 3.0 Amperes, thus showing that the resistances in the two paths in the superconducting state are equal and zero. The overshoot at Tc is most probably due to path B going superconducting a little earlier than A due to its location being lower than that of A and thus being exposed to the cooling gas earlier. **(b)** The dashed line shows the behavior with copper substituted for NbTi.

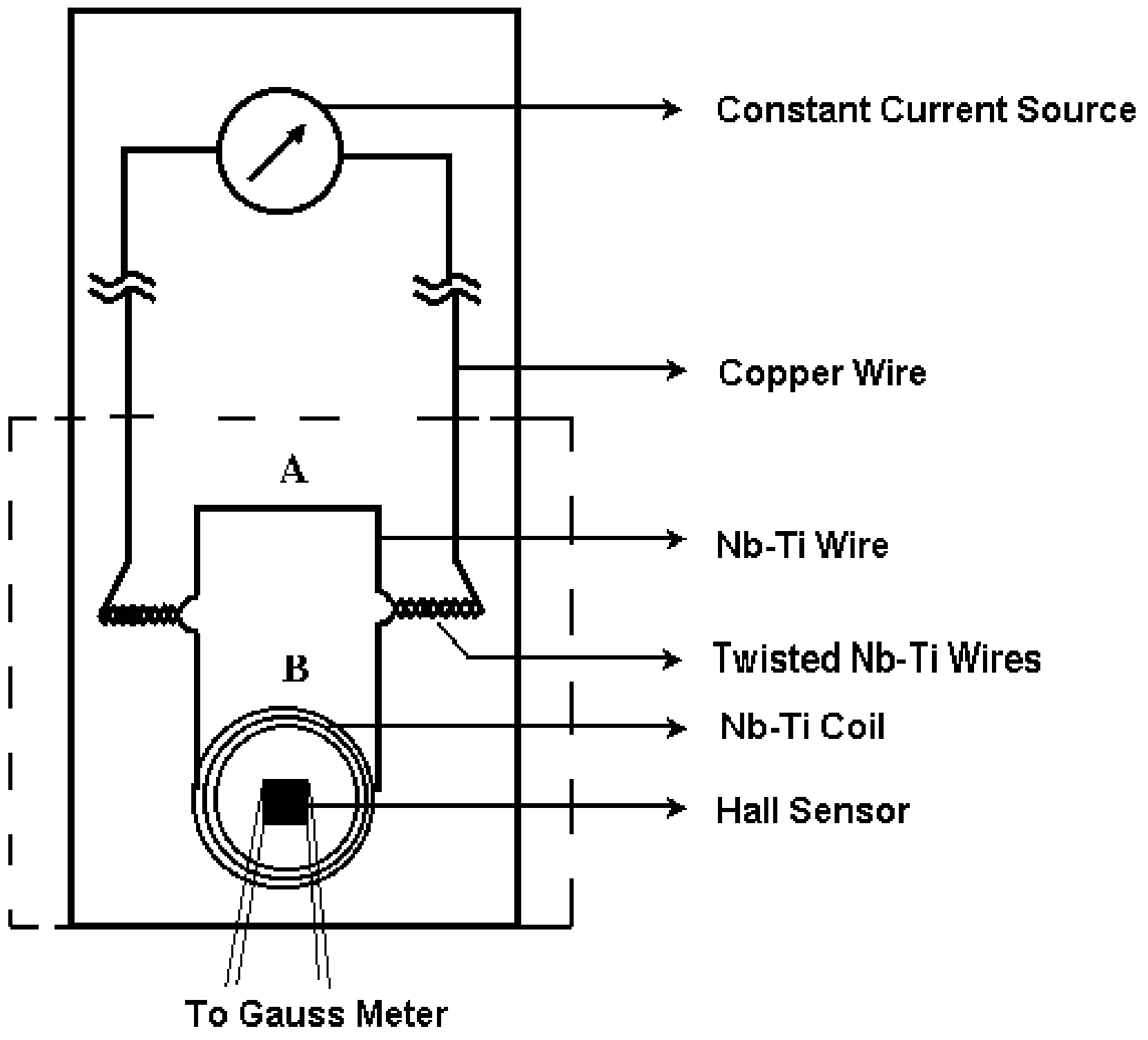

**Figure 1: Sarangi et al.,**

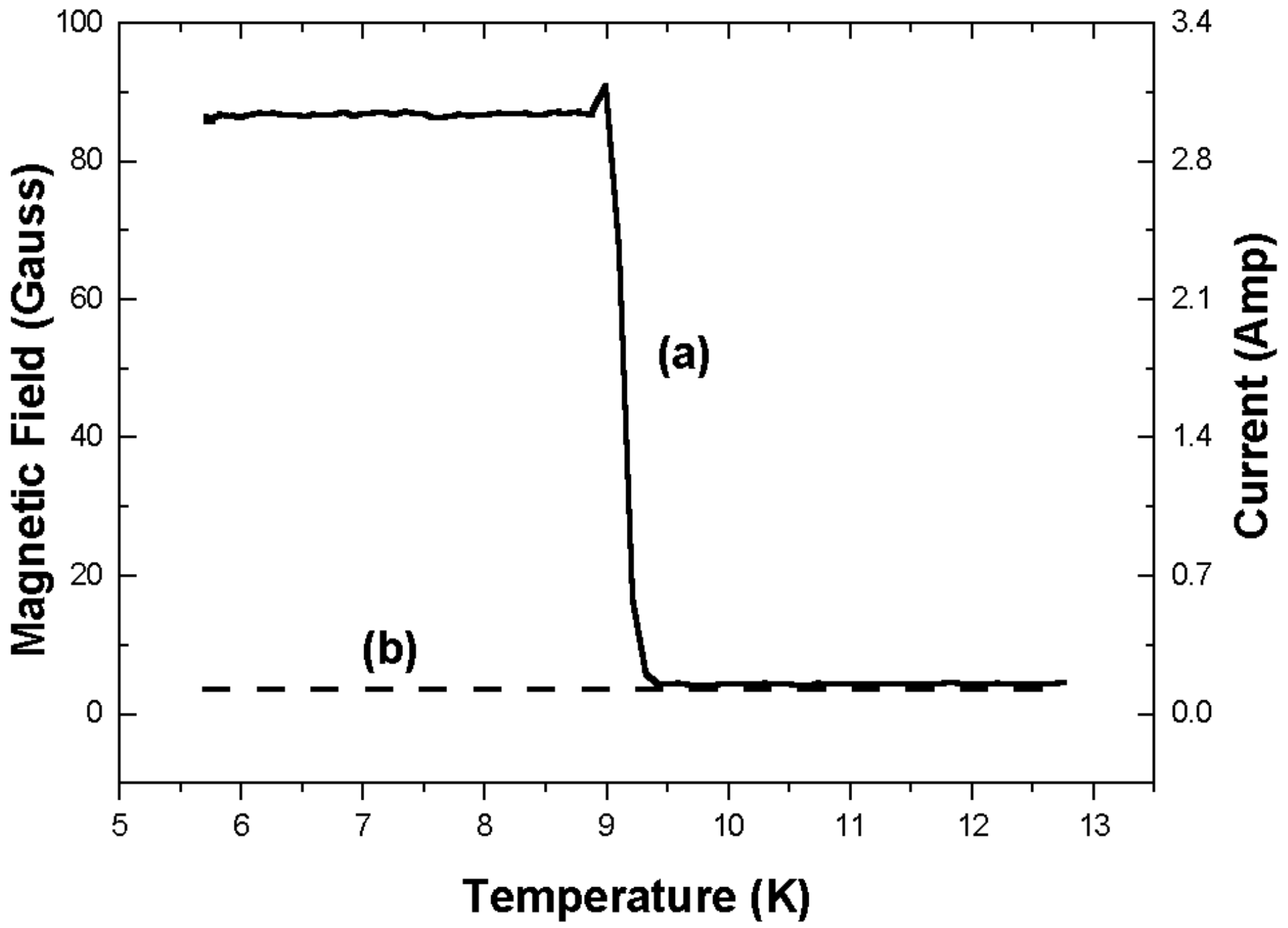

**Figure 2: Sarangi et al.,**

# Supplementary Material:

The niobium-titanium, NbTi wire was of gauge 22. For temperature variation an Oxford Instruments CF1300 continuous flow cryostat with an ITC 502 temperature regulator ( settability and stability of temperature: ± 0.1 K ) was used. The temperature was also independently monitored with a TRI research ruthenium oxide sensor. The magnetic field was measured using a Cryomagnetics model HSP-A Hall probe together with a Cryomagnetics model GM-700 Gauss meter. The Hall probe was calibrated at five different temperatures from room temperature to 7 K and the temperature independence and linearity with magnetic field were established as shown in figure S1.

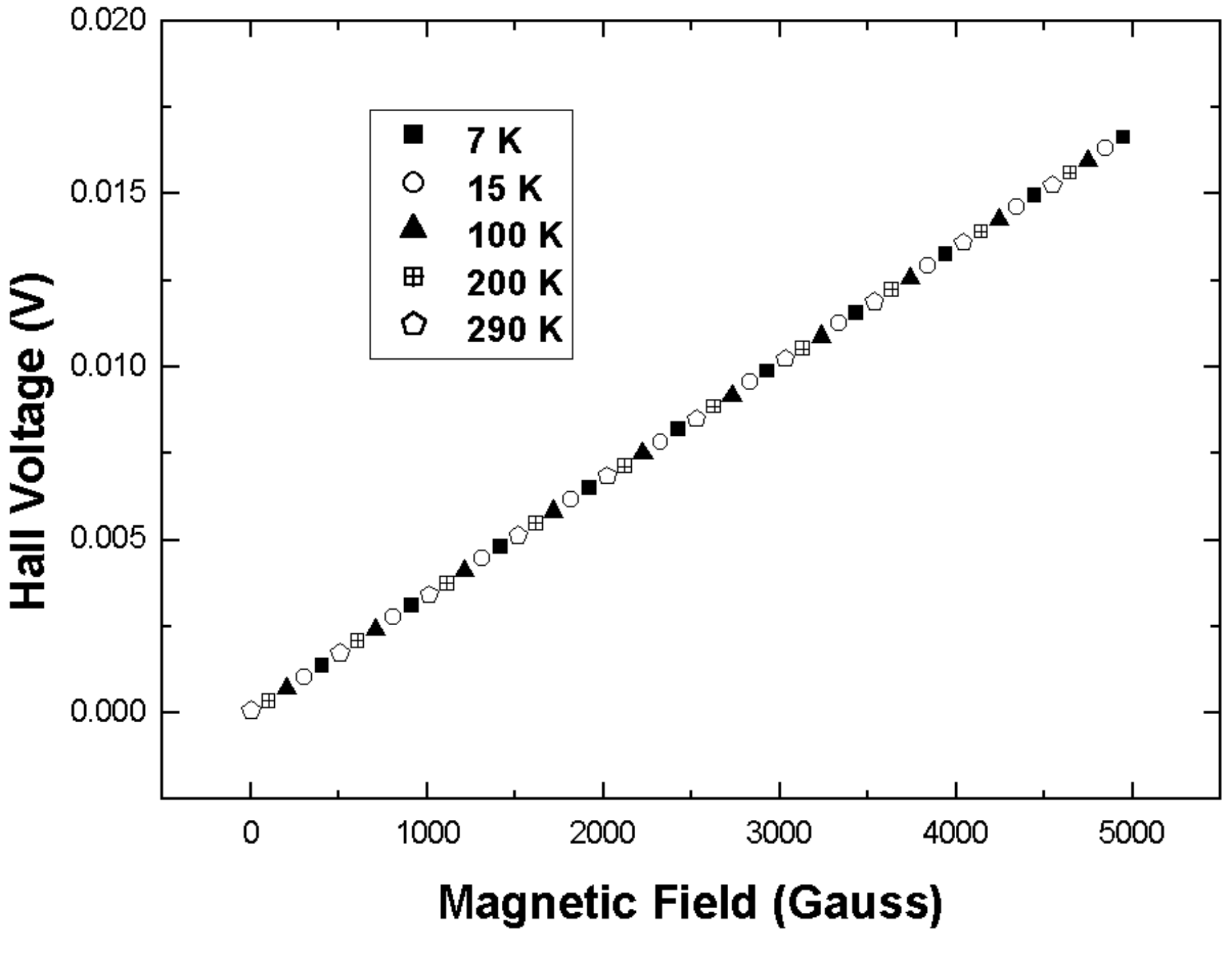

Fig. S1: The Hall voltage measured inside the cryostat at five different temperatures. An electromagnet was used to apply fields up to 5000 Gausss. In the figure only a few data points corresponding to each temperature are shown for the sake of clarity.